\documentclass[prd,preprint,nofootinbib,superscriptaddress]{revtex4}
\usepackage{amsmath,amssymb}
\usepackage{cancel}
\usepackage{graphicx}
\usepackage{lscape}

\newcommand{\lag}{\ensuremath{\mathcal{L}}}
\newcommand{\Op}{\ensuremath{\mathcal{O}}}
\newcommand{\PP}{\ensuremath{\mathcal{P}}}

\begin{document}

\title{The Effective Lagrangian for Bulk Fermions}
\date{\today}
\preprint{CAFPE-164/11}
\preprint{UGFT-294/11}

\author{Adri\'an Carmona}
\author{Jos\'e Santiago}
\affiliation{
CAFPE and Departamento de F\'{\i}sica Te\'orica y del 
Cosmos, \\
Universidad de Granada, E-18071 Granada, Spain}

\begin{abstract}
We compute the dimension 6 effective Lagrangian arising
from the tree level integration of an arbitrary number of 
bulk fermions in models with warped extra dimensions.
The coefficients of the effective operators are written in terms of
simple integrals of the metric and 
are valid for arbitrary warp factors, with or without an infrared
brane, and for a general Higgs profile. 
All relevant tree level fermion effects in electroweak and flavor
observables can be computed using this effective Lagrangian. 
\end{abstract}

\pacs{}

\maketitle

\section{Introduction}

Models with warped extra dimensions~\cite{Randall:1999ee} 
offer a calculable path to the study of electroweak symmetry
breaking (EWSB) induced by a strongly coupled
sector~\cite{ArkaniHamed:2000ds}. 
Even if EWSB does not proceed precisely through the dual of
the original Randall-Sundrum model, it is plausible that 
models with arbitrary warp factors provide us with a general enough
sampling of the parameter space of strong EWSB models to give us
the insight needed to decifer the LHC data.
It is therefore crucial to develop techniques that allow us to study
the phenomenological implications of models with warped extra
dimensions for arbitrary backgrounds. 

Indirect constraints on models with warped extra dimensions have been
extensively studied in the past (see for
instance~\cite{Huber:2001gw,Agashe:2004rs,Cabrer:2010si}). 
Most of these studies are
performed using a fixed background, and the few cases in which the
low energy effective Lagrangian is computed for an arbitrary
background, only the contribution of bulk gauge bosons has been taken
into account. Fermion contributions on the other hand
have been computed only for a fixed
background, typically for a finite number of KK modes and always for
specific models (using general results from intregration of
four-dimensional vector-like
fermions~\cite{delAguila:2000aa,delAguila:2008pw}, see also
\cite{Buras:2009ka}). The purpose of this work is to fill this gap by
computing the dimension 6 effective Lagrangian 
generated after the integration at tree
level of an arbitrary number of bulk fermions, including the effect of
the full tower. The result is written in terms of integrals of the
metric and the Higgs profile, thus being applicable to models with
arbitrary warp factor and Higgs localization.
EWSB effects can be included to all orders essentially only when
they enter directly or effectively as boundary
terms. In order to
consider a more general case, we assume a light Standard Model
(SM)-like Higgs with an
arbitrary profile. This is a very good approximation in general if
there is a light Higgs in the spectrum. For
instance in composite Higgs models, bounds on the S parameter typically
imply suppressed non-linear Higgs terms~\cite{Agashe:2004rs}.
In order to obtain the background independent results, we have used 5D
propagators to integrate out the full tower of bulk
fermions (these techniques can be also useful for 5D loop
calculations, see for instance~\cite{Csaki:2010aj}). 
Holographic methods would give the same results
(see~\cite{Davoudiasl:2009cd} for a detailed comparision of both methods). 

The rest of the article is organized as follows. We introduce our
notation and discuss how to integrate out bulk fermions using 5D
propagators in Section~\ref{integrating:out}. The different terms
entering the coefficients of the effective Lagrangian are computed in
terms of integrals of the metric and Higgs profiles in
Section~\ref{coefficients}. An application of the formalism developed
in these two sections is discussed in Section~\ref{example}, in
which we compute the corrections to SM Yukawa couplings induced by
bulk fermions. We conclude in Section~\ref{conclusions}. Technical
details on the calculation and results of the 5D propagators and the
general expression for the SM gauge and Yukawa couplings derived from
the effective Lagrangian are given in two appendices.

\section{Integrating out bulk fermions\label{integrating:out}}

In this section we will compute the effective Lagrangian
resulting from the tree level integration of an arbitrary number of
bulk fermions in models with warped
extra dimensions.
The only relevant assumption we make is to have a
light Higgs  
in the spectrum, responsible for electroweak symmetry breaking, 
but otherwise remain completely general.
Our methods can be 
generalized to include the effects of EWSB to all orders by including
the Higgs vev as part of our general background or through boundary
conditions (this is essential
for instance in Higgsless models~\cite{Csaki:2003dt}). For clarity, we
prefer to consider EWSB perturbatively as non-linear effects in the
Higgs vev do not interfere with the effects of bulk fermions that we
are interested in at the level of dimension 6 operators.
Also, we will focus in this work on the quark sector but the results
presented here are applicable, with minimal changes to the leptonic
sector too (see~\cite{delAguila:2008pw}).
With our assumption on EWSB, all fields can be classified according to
their SM quantum numbers. In this case, the only new quarks that
contribute to the effective Lagrangian at tree level and dimension 6
are the ones with the quantum numbers shown in table~\ref{multiplets}.
%%%%%%%%%%%%%%%%%%%%%%%%%%%%%%%%%%%%%%%%
\begin{table}[!ht]
{\scriptsize
\begin{tabular}{|c|ccccccc|}
\hline
$Q^{(m)}$ & $U$ & $D$ & $ \left(\begin{array}{c} U\\D
\end{array}\right)$ &$ \left(\begin{array}{c} X\\U
\end{array}\right)$ &$ \left(\begin{array}{c} D\\Y
\end{array}\right)$ & $ \left(\begin{array}{c} X\\U\\D
\end{array}\right)$ & $\left(\begin{array}{c} U\\D\\Y
\end{array}\right)$ 
\\ 
Notation & & & $Q$ 
& $Q^X$ & $Q^Y$ 
& $T^X$ & $T^Y$ 
\\
\hline
$SU(2)_L\times U(1)_Y$&
  $0_\frac{2}{3}$&$0_{-\frac{1}{3}}$&$2_\frac{1}{6}$&$2_\frac{7}{6}$
&$2_{-\frac{5}{6}}$
&$3_\frac{2}{3}$& $3_{-\frac{1}{3}}$
\\
\hline
\end{tabular}\vspace{0.5cm}
}
\caption{New quark multiplets,
  $Q^{(m)}$, that can mix with the SM fermions
through Yukawa 
couplings. The index $m$ labels the differen types of fermion multiplet
additions in the given order.
The electric charge is the sum of the third component of
isospin $T_3$ and the hypercharge $Y$. } \label{multiplets}
\end{table}
%%%%%%%%%%%%%%%%%%%%%%%%%%%%%%%%%%%%%%%%%

We consider a 5D space with a general warped metric
\begin{equation}
ds^2= a^2(z) \big[ \eta_{\mu\nu} dx^\mu \, dx^\nu - dz^2 \big],
\end{equation}
where $x^\mu$ denote the standard four extended space-time dimensions
and
$z$ parameterizes the extra dimension, which has two boundaries, 
$L_0 \leq z \leq L_1 $, called the UV and IR branes, respectively. The
warp factor is general but normalized to $a(L_0)=1$.
Popular choices for this warp factor are $a(z)=1$ for flat space and 
$a(z)=L_0/z$ for AdS$_5$. Also, the case of no IR brane can be simply
obtained by taking $L_1 \to \infty$.

Let us consider a number of bulk fermions, $\Psi^b(x,z)$, classified
according to their SM quantum numbers. Bulk fermions with quantum
numbers other than the ones in table~\ref{multiplets} do not
contribute to the effective Lagrangian we are computing and are
disregarded.  
We separate them into
their zero mode (when it exists) and non-zero mode components
\begin{equation}
\Psi^b_{L,R}(x,z)= f^{(0)b}_{iL,R}(z) \psi^{(0)i}_{L,R}(x)+ 
\tilde{\Psi}^b_{L,R} (x,z), \label{zeromodeseparation}
\end{equation}
where, from now on, repeated indices imply summation unless otherwise
stated, $i=1,2,3$ denotes the SM fermion generation index, 
and we have explicitly written the (4D) chiralities 
\begin{equation}
\Psi_{L,R} = \mathcal{P}_{L,R} \Psi, \quad
\mathcal{P}_{L,R} \equiv \frac{1\mp \gamma^5}{2}.
\end{equation}
The action can then be written
\begin{equation}
S_5=S_{SM}+\Delta S,\label{Stot}
\end{equation}
where the part of the action involving only zero modes gives, after
integration over the extra dimension, the 
SM action $S_{SM}$, for which we take for following convention,
\begin{equation}
S_{SM} = \int d^4x \Big\{ 
\bar{q}^i_L i\cancel{D} q^i_L
+\bar{u}^i_R i\cancel{D} u^i_R
+\bar{d}^i_R i\cancel{D} d^i_R
-\big(V^\dagger_{ij}\lambda^u_j \bar{q}^i_L \tilde{\phi} u_R^j
+ \lambda^d_i \bar{q}^i_L \phi d^i_R +\mathrm{h.c.}\big)
\Big\},
\label{SM:convention}
\end{equation}
whereas the part of the action involving heavy
modes can be written
\begin{equation}
\Delta S = \int d^4x \int_{L_0}^{L_1}dz\,
\Big\{
\overline{\tilde{\Psi}}^b\Big( \Op^{b}_K \delta^{bc} 
+ \Op_\varphi^{bc} \Phi^{bc}  \Big) \tilde{\Psi}^c
+\Big[
\overline{\tilde{\Psi}}^b J^b + \mathrm{h.c.} \Big]
\Big\},\label{DeltaS:generic}
\end{equation}
where $\Op_K$ represents the kinetic (including the SM gauge bosons
through the covariant derivative) 
and mass contributions and
$\Op_\varphi$ the couplings to the SM Higgs. $\Phi^{bc}$ represents
here the appropriate expression of the SM Higgs, according to the
quantum numbers of the fermions involved. Explicitly we have
\begin{eqnarray}
\overline{[2_Y]}[1_{Y+\frac{1}{2}}] &\to& \Phi\equiv \tilde{\phi},
\quad
\overline{[3_Y]}[2_{Y+\frac{1}{2}}] \to \Phi\equiv \phi^\dagger \frac{\sigma^a}{2},
\nonumber 
\\
\overline{[2_Y]}[1_{Y-\frac{1}{2}}] &\to& \Phi\equiv \phi,
\quad
\overline{[3_Y]}[2_{Y-\frac{1}{2}}] \to \Phi\equiv 
\tilde{\phi}^\dagger \frac{\sigma^a}{2},
\label{higgs:couplings}
\end{eqnarray}
where we have denoted in squared brackets the quantum numbers of the
two fermionic fields involved in the coupling,  
$\tilde{\phi}\equiv i \sigma_2 \phi^\ast$ denotes the
$Y=-1/2$ Higgs doublet and the Pauli matrices are in the basis of
$T_3^L=\pm 1,0$, as are the corresponding quark triplets.
Finally, the currents $J$, which  
contain a SM fermion and the SM Higgs, have the
generic form
\begin{equation}
J^b = J^b_L+J^b_R=  \Op^{bc}_{\varphi} \Phi^{bc}\big[ 
f^{(0)c}_{sL} \psi^{(0)s}_L
+ f^{(0)c}_{sR} \psi^{(0)s}_R \big].
\end{equation}
Due to chirality of the SM spectrum, only one of the two
components is non-vanishing for each type of heavy fermion.
In particular, using the notation in Table~\ref{multiplets}, the
non-vanishing currents can be written as
\begin{eqnarray}
J^U_L &=&
-\lambda_{U q_L^j}(z) V_{ji}  \tilde{\phi}^\dagger q^i_L,
\qquad
\qquad
\qquad
J^D_L =
-\lambda_{D q_L^i}(z)  \phi^\dagger q^i_L,
\\
J^Q_R &=&
-\lambda_{Q u_R^i}(z)  \tilde{\phi} u^i_R
-\lambda_{Q d_R^i}(z)  \phi d^i_R,
\\
J^{Q^X}_R &=&
-\lambda_{Q^X u_R^i}(z)  \phi u^i_R,
\qquad
\qquad
\qquad
J^{Q^Y}_R =
-\lambda_{Q^Y d_R^i}(z)  \tilde{\phi} d^i_R,
\\
J^{T^X}_L &=&
-\lambda_{T^x q_L^j}(z) V_{ji}\frac{\sigma^a}{2} \tilde{\phi}^\dagger q^i_L,
\qquad
\qquad
J^{T^Y}_L =
-\lambda_{T^Y q_L^j}(z) V_{ji} \frac{\sigma^a}{2} \phi^\dagger q^i_L,
\end{eqnarray} 
where the $\lambda_{\Psi\psi}(z)$ encode the extra dimensional
dependence of the effective Yukawa couplings.
All these operators are functions of $z$ that encode the dependence on
the metric and the wave functions of the different fields. Their explicit
expressions will be given in the next section.
Their mass dimensions are $[\Op_K]=1$, $[\Op_\varphi]=0$, $[J]=3$
(also $[\Psi]=2$).

We can integrate out the heavy fields at tree level 
by solving their classical equations of
motion 
\begin{equation}
\Big[\Op_K^b \delta^{bc} + \Op_\varphi^{bc} \Phi^{bc}\Big]\tilde{\Psi}^c = - J^b,
\end{equation}
and inserting the solution back in the action
\begin{equation}
\Delta S=
\int d^4xdz\,
\bar{J}^{b}(x,z) \tilde{\Psi}^b(x,z)
=\int \frac{d^4p}{(2\pi)^4} dz \bar{J}^a(p,z) \tilde{\Psi}^b(p,z),
\end{equation}
where in the last equality
we have switched to mixed position/momentum
space~\cite{ArkaniHamed:2001mi} by Fourier
transforming with respect to the four extended dimensions. We have
implicitly denoted the functions and their Fourier transforms by their
argument. 
$\tilde{\Psi}^b$ is supposed to be replaced with
the solution of the equation of motion which, in mixed
position/momentum space reads
\begin{equation}
\Op_{K}^b(p,z)
\tilde{\Psi}^b(p,z)
=-J^b(p,z)
-\int \frac{d^4p_1}{(2\pi)^4}
\Op^{bc}_{\phi}(z)  \Phi^{bc}(p_1)\tilde{\Psi}^c(p-p_1,z).
\end{equation}
Strictly speaking, one has to include the terms in $\Op_K$ containing
the SM gauge bosons as a convolution on the right hand side. Gauge invariance
guarantees however that we can forget about those terms and recover
them at the end by turning normal derivatives to covariant ones. We
have explicitly checked that both methods give the same result.
Let us define the Green's function of $\Op_{K}$ with the zero mode
subtracted 
\begin{equation}
\Op_{K}^{b}(p,z) P^{bc}_p(z,z^\prime) = \delta^{bc} \delta(z-z^\prime),
\quad \tilde{P}^{bc} = P^{bc} - P_{\mathrm{zero~mode}}^{bc},
\label{Greens:definition}
\end{equation}
which can be decomposed in their chiral (from the 4D
point of view) components~\cite{Carena:2004zn}
\begin{equation}
P_{LL}=\PP_L P \PP_R ,\quad P_{LR}=\PP_L P \PP_L,
\quad P_{RR}=\PP_R P \PP_L ,\quad  P_{RL}=\PP_R P \PP_R.
\label{P:components}
\end{equation}
The different components can be written as follows
\begin{eqnarray}
\tilde{P}^{ab}_{LL}(p;z,z^\prime) &=& 
(\cancel{p}L_1) \hat{P}^{ab}_{LL}(p;z,z^\prime),\quad
\tilde{P}^{ab}_{RR}(p;z,z^\prime) = 
(\cancel{p}L_1) \hat{P}^{ab}_{RR}(p;z,z^\prime), \nonumber \\
\tilde{P}^{ab}_{LR}(p;z,z^\prime) &=& 
\hat{P}^{ab}_{LR}(p;z,z^\prime),\qquad \quad
\tilde{P}^{ab}_{RL}(p;z,z^\prime) = 
\hat{P}^{ab}_{RL}(p;z,z^\prime),\label{hatted:def}
\end{eqnarray} 
with $\hat{P}(p;z,z^\prime)= \hat{P}^{(0)}(z,z^\prime) + (p^2L_1^2)
\hat{P}^{(1)}(z,z^\prime) + \ldots$,
where we have introduced the appropriate powers of $L_1$ to keep all
components of the Green's functions with the same (vanishing) mass
dimension.  
The solution of the equation of motion can be written in terms of
these Green's functions by iteration
\begin{eqnarray}
\tilde{\Psi}^b(p,z) 
&=&
-\int dz^\prime \tilde{P}^{bc}_p(z,z^\prime)
\bigg\{
J^c(p,z^\prime)
+
\int \frac{d^4p_1}{(2\pi)^4}
\Op^{cd}_{\phi}(z^\prime)  \Phi^{cd}(p_1)
\tilde{\Psi}^d(p-p_1,z^\prime)
\bigg\} \\
&=&
-\int dz^\prime \tilde{P}^{bc}_p(z,z^\prime)
\bigg\{
J^c(p,z^\prime)
\nonumber \\
&&
\phantom{-\int dz^\prime \tilde{P}^{bc}_p(z,z^\prime)}
-\int dz^{\prime\prime}
\int \frac{d^4p_1}{(2\pi)^4}
\Op^{cd}_{\phi}(z^\prime)  \Phi^{cd}(p_1)
\tilde{P}^{de}_{p-p_1}(z^\prime,z^{\prime\prime})
J^e(p-p_1,z^{\prime\prime})
\bigg\}
+\ldots\,.\nonumber
\end{eqnarray}
Inserting this solution back in the action and taking into account
the following properties
\begin{eqnarray}
&&P^{bc}\neq 0 \Rightarrow \bar{J}^{b}_L J^c_R = \bar{J}^{b}_R J^c_L
  =0,
\\
&&
\Op^{bc}_\phi \neq 0 \Rightarrow
\bar{J}^b_L J^c_L=\bar{J}^b_R J^c_R=0,
\end{eqnarray}
we get the following effective Lagrangian
\begin{eqnarray}
\mathcal{L}_6&=&
-L_1 
\int dzdz^\prime 
\left\{
\bar{J}_L^b(x,z) \hat{P}^{(0)bc}_{RR}(z,z^\prime) 
i\cancel{D} J^c_L(x,z^\prime)
+
(L \leftrightarrow R)
\right\}
\nonumber \\
&+&
\int dz dz^\prime dz^{\prime\prime}
\Big\{
\bar{J}_L^b(x,z) \hat{P}^{(0)bc}_{RL}(z,z^\prime)
\Op^{cd}_{\phi}(z^\prime) \Phi^{cd}(x)
\hat{P}^{(0)de}_{RL}(z^\prime,z^{\prime\prime})
J^e_R(x,z^{\prime\prime})
+\mathrm{h.c.}
\Big\},\label{L6:first}
\end{eqnarray}
where we have switched back to position space (including the terms
coming from the SM gauge fields), factored out the integration over 4D
space to obtain the effective Lagrangian and the terms that we have not
written are either absent due to the chirality of the SM spectrum or
give rise to operators of dimension higher than 6.

In order to write our effective Lagrangian in the standard basis
of~\cite{Buchmuller:1985jz} we need to manipulate the operators in
Eq.~(\ref{L6:first}). After applying the Leibniz rule for
the covariant derivative, using the equations of motion from
$\lag_{SM}$ for the SM fermions and Fierz reordering, we end up with
the following effective Lagrangian~\cite{delAguila:2000aa}
\begin{eqnarray}
\lag_{\mathrm{eff}} &=&
\lag_{\mathrm{SM}} +
\frac{1}{\Lambda^2}
\Big[
\sum_{\psi_L} (\alpha^{(3)}_{\phi \psi_L})_{ij}
 (\Op^{(3)}_{\phi\psi_L})^{ij}
+\sum_{\psi} (\alpha^{(1)}_{\phi \psi})_{ij} (\Op^{(1)}_{\phi\psi})^{ij}
\nonumber \\
&&
\phantom{\lag_{\mathrm{SM}}+
\frac{1}{\Lambda^2}
\Big[}
+ (\alpha_{\phi \phi})_{ij} (\Op_{\phi \phi})^{ij}
+\sum_{\psi_R} (\alpha_{\psi_R \phi})_{ij} (\Op_{\psi_R \phi})^{ij}
+ \mathrm{h.c.} \Big], \label{Leff}
\end{eqnarray}
where we have explicitely factored out two powers of the cut-off 
$\Lambda\sim L_1^{-1}$ (so that the coefficients of the different
operators are dimensionless), 
the sums run over all left-handed (LH) SM fields $\psi_L$, all SM fields
$\psi$ and all SM right-handed (RH) fields $\psi_R$, respectively. The different
operators are defined as follows
\begin{eqnarray}
(\Op^{(3)}_{\phi\psi_L})^{ij} &=& (\phi^\dagger \sigma^I i D_\mu \phi)
  (\bar{\psi}^i_L \sigma^I \gamma^\mu \psi_L^j),
\\
(\Op^{(1)}_{\phi\psi})^{ij} &=& (\phi^\dagger  i D_\mu \phi)
  (\bar{\psi}^i \gamma^\mu \psi^j),
\\
(\Op_{\phi \phi})^{ij} &=& (\phi^{\mathrm{T}} \epsilon i D_\mu  \phi)
  (\bar{u}^i_R \gamma^\mu d_R^j),
\\
(\Op_{\psi_R \phi})^{ij} &=& (\phi^\dagger  \phi)
  (\bar{\psi}^i_L \Phi_{\psi_R} \psi_R^j),
\end{eqnarray}
where $\Phi_{u_R}=\tilde{\phi}$ and $\Phi_{d_R}=\phi$.
The first three kinds of operators come from the first line in
Eq.~(\ref{L6:first}) and involve contributions from one kind of
multiplet in table~\ref{multiplets} at a time. The last operator, on
the other hand, receives corrections from both lines in 
Eq.~(\ref{L6:first}) and therefore involves one or 
two different heavy multiplets at a time.
The resulting coefficients of the effective operators are collected in
tables~\ref{tcoefficients} and \ref{mixedcoefficients}, written in
terms of the 
following overlap integrals
\begin{eqnarray}
\beta^{\Psi;\chi}_{\psi^i \psi^{\prime j}} &\equiv&
-L_1 \int dz dz^\prime \,
\lambda^{\dagger}_{\psi^i \Psi^b}(z)
\hat{P}^{(0)\Psi^b \Psi^c}_{\chi \chi}(z,z^\prime) 
\lambda_{\Psi^c \psi^j}(z^\prime), \label{beta:coeffs}
\\
\gamma^{\Psi \Psi^\prime; \chi \chi^\prime}_{\psi^i \psi^{\prime j}} &\equiv&
\int dz dz^\prime dz^{\prime\prime}\,
\lambda^{\dagger}_{\psi^i \Psi^b}(z)
\hat{P}^{(0)\Psi^b \Psi^c}_{\chi \chi^\prime}(z,z^\prime) 
\Op_{\varphi}^{\Psi^c \Psi^{\prime d}}(z^\prime)
\hat{P}^{(0)\Psi^{\prime d} \Psi^{\prime e}}_{  \chi\chi^\prime
}(z^\prime,z^{\prime\prime}) 
\lambda_{\Psi^{\prime e} \psi^{\prime j}}(z^{\prime \prime}).
\label{gamma:coeffs}
\end{eqnarray}
The only remaining pieces to compute the effective Lagrangian are the
explicit expressions of the $\Op_\varphi(z)$ and $\lambda(z)$
functions and the zero momentum propagators. The former two are done in
the next section whereas the latter are given in Appendix~\ref{green:App}.
%%%%%%%%%%%%%%%%%%%%%%%%%%%%%%%%%%%%%%%%%%%%%
\begin{table}[ht]
\caption{Coefficients $\alpha_x^m$ 
resulting from the
integration of an arbitrary number of each type of bulk
quark. The coefficients $\beta^{\Psi;\chi}_{\psi^i \psi^{\prime j}}$
are defined in Eq.(\ref{beta:coeffs}).}\label{tcoefficients}
\footnotesize
\begin{tabular}{c|ccccccc}
$Q^{(m)}$ \rule[-.5cm]{0cm}{1cm} & 
$\frac{(\alpha^{(1)}_{\phi q})_{ij}}{\Lambda^2}$  
& $\frac{(\alpha^{(3)}_{\phi q})_{ij}}{\Lambda^2}$ &
$\frac{(\alpha_{\phi u})_{ij}}{\Lambda^2}$ & 
$\frac{(\alpha_{\phi d})_{ij}}{\Lambda^2}$ 
&$\frac{(\alpha_{\phi \phi})_{ij}}{\Lambda^2}$ &
$\frac{(\alpha_{u \phi})_{ij}}{\Lambda^2}$& 
$\frac{(\alpha_{d \phi})_{ij}}{\Lambda^2}$    
\\ \hline
$U$ \rule{0cm}{.8cm}
&$ 
\frac{1}{4} V^\dagger_{ik} 
\beta^{U;R}_{q^k_L q^l_L}
V_{lj}
$ 
&$-\frac{(\alpha^{(1)}_{\phi q})_{ij}}{\Lambda^2}$  & $-$ & $-$ & $-$ &
$2\frac{(\alpha^{(1)}_{\phi q})_{ik}}{\Lambda^2} V^\dagger_{kj} 
\lambda^u_j$ &
$-$   
\\
$ D$ \rule{0cm}{.8cm}
&$
-\frac{1}{4} 
\beta^{D;R}_{q^i_L q^j_L}
$
&$\frac{(\alpha^{(1)}_{\phi q})_{ij}}{\Lambda^2}$  & $-$ & $-$ & $-$ &
$-$& $-2\frac{(\alpha^{(1)}_{\phi q})_{ij}}{\Lambda^2} \lambda^d_j$ 
\\
$Q$  \rule{0cm}{.8cm}
& $-$ & $-$ &
$ -\frac{1}{2} \beta^{Q;L}_{u^i_R u^j_R}$
&$ \frac{1}{2} \beta^{Q;L}_{d^i_R d^j_R}$
&$ -\beta^{Q;L}_{u^i_R d^j_R}$
&$-V^\dagger_{ik}\lambda^u_k\frac{(\alpha_{\phi u})_{kj}}{\Lambda^2}$
&$\lambda^d_i\frac{(\alpha_{\phi d})_{ij}}{\Lambda^2}$
\\
 $Q^X$  \rule{0cm}{.8cm}
& $-$ & $-$ &
$ \frac{1}{2} \beta^{Q^X;L}_{u^i_R u^j_R}$
&$ -$
&$ -$
&$V^\dagger_{ik}\lambda^u_k\frac{(\alpha_{\phi u})_{kj}}{\Lambda^2}$
&$-$
\\
 $Q^Y$  \rule{0cm}{.8cm}
& $-$ & $-$ &
$ -$
&$-\frac{1}{2} \beta^{Q^Y;L}_{d^i_R d^j_R}$
&$ -$
&$-$
&$-\lambda^d_i\frac{(\alpha_{\phi d})_{ij}}{\Lambda^2}$
\\
 $T^X$  \rule{0cm}{1cm}
&$\frac{3}{16} V^\dagger_{ik} \beta^{T^X;R}_{q^k_L q^l_L} V_{lj} $
&$\frac{1}{3}
\frac{(\alpha^{(1)}_{\phi q})_{ij}}{\Lambda^2}$  & $-$ & $-$ & $-$ &
$\frac{2}{3}
\frac{(\alpha^{(1)}_{\phi q})_{ik}}{\Lambda^2} V^\dagger_{kj} \lambda^u_j$ &
$\frac{4}{3}
\frac{(\alpha^{(1)}_{\phi q})_{ij}}{\Lambda^2} \lambda^d_j$    
\\
 $T^Y$  \rule{0cm}{1cm}
&$-\frac{3}{16} V^\dagger_{ik} \beta^{T^Y;R}_{q^k_L q^l_L} V_{lj} $
&$-\frac{1}{3}
\frac{(\alpha^{(1)}_{\phi q})_{ij}}{\Lambda^2}$  & $-$ & $-$ & $-$ &
$-\frac{4}{3}
\frac{(\alpha^{(1)}_{\phi q})_{ik}}{\Lambda^2} V^\dagger_{kj} 
\lambda^u_j$ &
$-\frac{2}{3}
\frac{(\alpha^{(1)}_{\phi q})_{ij}}{\Lambda^2} \lambda^d_j$    
\end{tabular}
\end{table} 
%%%%%%%%%%%%%%%%%%%%%%%%%%%%%%%%%%%%%%%%%%%%%%%%%%%%%%%%%%%%%%%%%
\begin{table}[!hb]
\caption{Coefficients $\alpha_x^{mn}$ resulting from 
mixing between bulk multiplets.
The coefficients
$\gamma^{\Psi \Psi^\prime; \chi \chi^\prime}_{\psi^i \psi^{\prime j}}$
 are defined in Eq. (\ref{gamma:coeffs}).
\label{mixedcoefficients}}
\scriptsize
\begin{center}
\begin{tabular}{c||c|c|c|c|c|c|c|c}
&
$U, \; Q$ &
$U, \; Q^X$ &
$D, \; Q$ &
$D, \;Q^Y$ &
$Q, \; T^X $ &
$Q, \; T^Y$ &
$Q^X, \; T^X $ &
$Q^Y, \; T^Y$ \\
\hline
\rule{0cm}{.8cm} 
$ \frac{(\alpha^{mn}_{u\phi})_{ij}}{\Lambda^2} $ & 
$V_{ik}^\dagger \gamma^{UQ;RL}_{q^k_L u^j_R}$ &
$V_{ik}^\dagger  \gamma^{UQ^X;RL}_{q^k_L u^j_R}$ &
$-$ &
$-$ &
$\frac{1}{4} V_{ik}^\dagger \gamma^{T^XQ;RL}_{q^k_L u^j_R} $ &
$\frac{1}{2} V_{ik}^\dagger \gamma^{T^YQ;RL}_{q^k_L u^j_R} $ &
$-\frac{1}{4} V_{ik}^\dagger \gamma^{T^XQ^X;RL}_{q^k_L u^j_R}$& 
$-$ \\
\rule{0cm}{.8cm} 
$ \frac{(\alpha^{mn}_{d\phi})_{ij}}{\Lambda^2}$ &
$-$ &
$-$ &
$\gamma^{DQ;RL}_{q^i_L d^j_R} $ &
$\gamma^{DQ^Y;RL}_{q^k_L d^j_R} $ &
$\frac{1}{2} V_{ik}^\dagger  \gamma^{T^XQ;RL}_{q^k_L d^j_R}$ &
$\frac{1}{4} V_{ik}^\dagger \gamma^{T^YQ;RL}_{q^k_L d^j_R}$ &
$-$ &
$-\frac{1}{4} V_{ik}^\dagger  \gamma^{T^YQ^Y;RL}_{q^k_L d^j_R}$
\end{tabular}
\end{center}
\vspace{0.5cm}
\end{table}
%%%%%%%%%%%%%%%%%%%%%%%%%%%%%%%%%%%%%%%%%%%%%%%%%%%%%%%%%%%%%%

\section{General expression for the effective
  coefficients\label{coefficients}} 

Let us now give the explicit form of the different terms entering the
coefficients of the effective Lagrangian.
The relevant part of the action for the bulk fermions is
given by
\begin{equation}
S=
\int d^4x dz\,
a^4 
\overline{\Psi}^b \Big\{ 
\Big[ i\cancel{D} + \Big( \partial_z + 2
  \frac{a^\prime}{a} \Big) \gamma^5 - a M_{\Psi^b}  \Big]
\delta^{bc} 
-a  f_h \lambda^{(5)}_{bc} \Phi^{bc}
\Big\}\Psi^c
+S_\mathrm{bound.},
\end{equation}
where $S_{\mathrm{bound.}}$ denotes any possible boundary term,
$b,c$ run over all bulk multiplets with quantum numbers
appearing in table~\ref{multiplets}, 
$D$ denotes the SM covariant derivative,
we have assumed a bulk
Higgs with a zero mode profile given by $f_h(z)$, normalized to
\begin{equation}
\int_{L_0}^{L_1} dz\, a^3 f_h^2 =1,
\end{equation} 
$\lambda^{(5)}_{bc}$
denote the 5D Yukawa couplings and
$\Phi^{bc}$ represents the correct combination of the SM Higgs
appropriate for the quantum numbers of $\Psi_{b,c}$ as described in
Eq.~(\ref{higgs:couplings}).
In the expressions above, we have assumed the Higgs to be a 5D
scalar. In models of Gauge-Higgs unification, the Higgs comes from a
gauge boson and the replacement $f_h \to a^{-1} f_h$ should be
made. Subtleties related to a boundary Higgs will be discussed in
section~\ref{example}. 
This action can be simplified with the following field redefinition 
\begin{equation}
\Psi^b \to a^{-2} a_{M_b}^{-\gamma^5/2} \Psi^b, \label{psi:redefinition}
\end{equation}
where we have defined the effective (mass dependent) metric
\begin{equation}
a_{M_b}(z) \equiv \exp \Big[ -2 \int_{L_0}^z \mathrm{d}z^\prime \,
  a(z^\prime) 
M_{\Psi_b}(z^\prime) \Big]
=a_{-M_b}^{-1}(z).
\end{equation}
After this field redefinition, the action can be written 
\begin{equation}
S=
\int d^4x dz\,
\overline{\Psi}^b \Big[ 
\Big( i\cancel{D} a_{M_b}^{-\gamma^5}  +  \gamma^5 \partial_z \Big)
 \delta^{bc} 
-a a_{M_b}^{\gamma^5/2} f_h \lambda^{(5)}_{bc} \Phi_{bc}
a_{M_c}^{-\gamma^5/2}
\Big]\Psi^c
+S_\mathrm{bound.}.
\end{equation}

We separate now the bulk fermions into
their zero mode and non-zero mode components as in
Eq.(\ref{zeromodeseparation}) 
\begin{equation}
\Psi^b_{L,R}(x,z)= f^{(0)b}_{iL,R} \psi^{(0)i}_{L,R}(x)+ 
\tilde{\Psi}^b_{L,R} (x,z).
\end{equation}
These zero modes realize the SM fermionic spectrum. We have denoted
with $i$ the flavor of
the canonically normalized fermion zero modes.
Note that we have allowed these fermion zero modes 
to be shared by several different
bulk fields, as happens when boundary conditions mix different bulk fields. 
In the basis after the redefinition (\ref{psi:redefinition}) the zero
mode profiles are
just constants fixed by the boundary conditions and
the orthonormality condition
\begin{eqnarray}
&&
f^{Q\dagger}_{q^i_L} \left[\int dz \, a_{M_Q}\right] f^Q_{q^j_L} =
f^{U\dagger}_{u^i_R} \left[\int dz \, a^{-1}_{M_U}\right] f^U_{u^j_R} =
f^{D\dagger}_{d^i_R} \left[\int dz \, a^{-1}_{M_D}\right] f^D_{d^j_R} =
\delta_{ij},\label{zero:modes:orthonormality}
\end{eqnarray}
where as usual, a sum over all possible values of $Q$, $U$ or $D$
is understood. 
We can further require that the zero mode profiles diagonalize the
zero mode Yukawa couplings for the down sector and the right part of
the Yukawa couplings for the up sector. This amounts to
\begin{eqnarray}
f^{Q\dagger}_{q^i_L} 
\left [\int dz a f_h
a_{\frac{M_Q}{2}}\lambda^{(5)}_{QU} 
a_{-\frac{M_U}{2}}
\right] f^U_{u^j_R} &=& V^\dagger_{ij} \lambda^u_j, 
\nonumber \\
f^{Q\dagger}_{q^i_L}  \left [\int dz a f_h
a_{\frac{M_Q}{2}}\lambda^{(5)}_{QD} 
a_{-\frac{M_D}{2}}
\right] f^D_{d^j_R} &=&\delta_{ij} \lambda^d_j.
\label{zero:modes:yukawas}
\end{eqnarray}
Note that for every zero mode profile that satisfy
Eq. (\ref{zero:modes:orthonormality}) we can define new ones
\begin{equation}
f^Q_{q^i_L} \to f^Q_{q^i_L} V^L_{ij},
\quad
f^U_{u^i_R} \to f^U_{u^i_R} V^{uR}_{ij},
\quad
f^D_{d^i_R} \to f^D_{d^i_R} V^{dR}_{ij},
\end{equation}
with $V^L, V^{uR}, V^{dR}$ $3\times 3$ unitary matrices so that the
new profiles satisfy both (\ref{zero:modes:orthonormality}) and
(\ref{zero:modes:yukawas}).
This separation allows us to write the action in the form of
Eq.~(\ref{Stot}), where 
Eqs.~(\ref{zero:modes:orthonormality}-\ref{zero:modes:yukawas})
guarantee that the SM action satisfies the convention in
Eq.~(\ref{SM:convention}). 
The quadratic operator reads
\begin{equation}
\Op_K^b =
\Big[ i\cancel{D} a_{M_b}^{-\gamma^5}  +  \gamma^5 \partial_z \Big]
+\mbox{ bound. terms},
\end{equation}
from which we can compute the relevant propagators as discussed in
Appendix~\ref{green:App}.
The Higgs operator is given by
\begin{equation}
\Op_\varphi^{bc} = 
-a a_{M_b}^{\gamma^5/2} f_h \lambda^{(5)}_{bc} 
a_{M_c}^{-\gamma^5/2}
+\mbox{ bound. terms}. \label{Ophi:explicit}
\end{equation}
Finally,
the explicit expressions for the non-vanishing currents read
\begin{eqnarray}
\lambda_{Uq^i_L}(z) &=&
\sum_Q
a f_h  a_{M_u}^{-\frac{1}{2}}
 \lambda_{UQ}^{(5)} a_{M_Q}^{\frac{1}{2}}f^Q_{q_L^j}V^\dagger_{ji},
\\
\lambda_{Dq^i_L}(z) &=&
\sum_Q
a f_h  a_{M_D}^{-\frac{1}{2}}
 \lambda_{DQ}^{(5)} a_{M_Q}^{\frac{1}{2}}f^Q_{q_L^i},
\\
\lambda_{Q u^i_R}(z)&=&
\sum_U
a f_h  a_{M_Q}^{\frac{1}{2}}
 \lambda_{QU}^{(5)} a_{M_U}^{-\frac{1}{2}}f^U_{u_R^i}, 
\\
\lambda_{Qd^i_R}(z)&=& 
\sum_D
a f_h  a_{M_Q}^{\frac{1}{2}}
 \lambda_{QD}^{(5)} a_{M_D}^{-\frac{1}{2}}f^D_{d_R^i},
\\
\lambda_{Q^Xu^i_R}(z) &=&
\sum_U
a f_h a_{M_{Q^X}}^{\frac{1}{2}}
 \lambda_{Q^X U}^{(5)} a_{M_U}^{-\frac{1}{2}}f^U_{u_R^i},
\\
\lambda_{Q^Yd^i_R}(z) &=&
\sum_D 
a f_h a_{M_{Q^Y}}^{\frac{1}{2}}
 \lambda_{Q^Y D}^{(5)} a_{M_D}^{-\frac{1}{2}}f^D_{d_R^i},
\\
\lambda_{T^X q^i_L} (z)&=&
\sum_Q 
a f_h a_{M_{T^X}}^{-\frac{1}{2}}
 \lambda_{T^x Q}^{(5)} a_{M_Q}^{\frac{1}{2}} f^Q_{q_L^j} V^\dagger_{ji},  
\\
\lambda_{T^Yq^i_L}(z) &=&
\sum_Q 
a f_h a_{M_{T^Y}}^{-\frac{1}{2}}
 \lambda_{T^Y Q}^{(5)} a_{M_Q}^{\frac{1}{2}}
 f^Q_{q_L^j}V^\dagger_{ji}, 
\label{lambdas:explicit}
\end{eqnarray} 
where possible contributions to boundary Yukawa couplings have not been
explicitly written. 

We can now plug these explicit expressions, 
Eqs.(\ref{Ophi:explicit}-\ref{lambdas:explicit}) 
and (\ref{P:explicit:first}-\ref{P:explicit:last}) in
Eqs. (\ref{beta:coeffs},\ref{gamma:coeffs}) to
compute the coefficients of the general effective Lagrangian for an
arbitrary warp factor and Higgs profile. 
As a example of the use of these general equations we will compute in
the next section the effect of bulk fermions in SM quark Yukawa
couplings.

\section{Application: Higgs couplings \label{example}}

As a straight-forward application of our formalism we compute
flavour violating Yukawa couplings in the down sector of 
a simplified model with two bulk fermions with the quantum numbers of
$Q$ and $D$ (following the notation in Table~\ref{multiplets}) and
boundary conditions $Q\sim[++]$, $D\sim[--]$. The two signs denote the
boundary condition at the UV and IR brane, respectively, and a $+$
($-$) denotes Dirichlet boundary conditions for the RH (LH) component
of the bulk field at the corresponding brane.
Their zero modes give rise to the SM $q_L$ and $d_R$ quarks. 
We want to compute the modified SM Yukawa couplings that
result after the integration of the heavy fields.
This has been studied recently in
Refs.~\cite{Azatov:2009na,Casagrande:2008hr}.  
In particular,
Ref.~\cite{Azatov:2009na} clarifies the implications of
brane-localized Yukawa 
couplings for fields with Dirichlet boundary conditions. As we will
see, the use of 5D propagators makes it apparent the ambiguity of
these couplings and a very simple regularization of the brane terms
gives directly the same result as previously obtained with other
methods.~\footnote{Similar ambiguities occur in brane localized
  kinetic terms for fields with Dirichlet boundary
  conditions~\cite{delAguila:2003bh}. In that case, there is a well-defined
  prescription to deal with such ambiguities in terms of field
  redefinitions and classical renormalization~\cite{delAguila:2006kj}.} 
The SM fermion gauge and Yukawa couplings derived from the effective
Lagrangian in Eq. (\ref{Leff}) were computed
in~\cite{delAguila:2000aa}. We collect the main
results in Appendix~\ref{trilinear:App}. The modified Yukawa
couplings can be written, in the physical basis (\textit{i.e.} 
with diagonal fermion
masses including effects of order $v^2/\Lambda^2$)
\begin{eqnarray}
\lag^H&=&-\frac{1}{\sqrt{2}}(
\bar{u}^i_L Y^{u}_{ij}
 u^j_R+
\bar{d}^i_L Y^{d}_{ij} d^j_R) H+ \mathrm{h.c.}, 
\end{eqnarray}
where the coefficients read
\begin{eqnarray}
Y^{u}_{ij}&=& \delta_{ij}\lambda^u_j-\frac{v^2}{\Lambda^2}
\left(V_{ik}(\alpha_{u \phi})_{kj} +
\frac{1}{4} \delta_{ij} [V_{ik}(\alpha_{u \phi})_{kj}+(\alpha_{u
\phi})^\dagger_{ik} V^\dagger_{kj}]\right)  ,\nonumber \\
Y^{d}_{ij}&=& \delta_{ij}\lambda^d_j-\frac{v^2}{\Lambda^2}
\left( (\alpha_{d \phi})_{ij}+\frac{1}{4}\delta_{ij} (\alpha_{d
\phi} +\alpha_{d \phi}^\dagger)_{ij} \right),
\end{eqnarray} 
with $v\approx 246$ GeV the Higgs vacuum expectation value.
Note that we have not written couplings proportional to $\partial_\mu
H$ because they vanish in our case due to the hermiticity of the
corresponding operator coefficients (see
Appendix~\ref{trilinear:App}).
These modified couplings are thus fully determined by two coefficients
that read, in our example,
\begin{eqnarray}
\frac{ V_{ik}(\alpha_{u\phi})_{kj}}{\Lambda^2}&=&
\frac{1}{2}\lambda^u_i \beta^{Q;L}_{u^i_R u^j_R},
\\
\frac{(\alpha_{d\phi})_{ij}}{\Lambda^2}&=&
\frac{1}{2}\beta^{D;R}_{q^i_L q^j_L} \lambda^u_j
+\frac{1}{2}\lambda^d_i \beta^{Q;L}_{d^i_R d^j_R}
+\gamma^{DQ;RL}_{q^i_L d^j_R}. \label{alphadphi}
\end{eqnarray}
Using the explicit expressions for the propagators in
Appendix~\ref{green:App}, we
can compute the corrections to the Yukawa couplings for arbitrary
backgrounds and Higgs profiles.
In the case of an IR boundary Higgs, the terms proportional to $\beta$
contain only Yukawa couplings of fields with Neumann boundary
conditions. They are well defined in the thin brane limit and a direct
application of our formulae reproduces the results in the
literature. The term proportional to $\gamma$ on the other hand,
involves a Yukawa coupling of fields with Dirichlet boundary
conditions and therefore should \textit{a priori} vanish. We will show
that the result is ambiguous in the thin brane limit but a simple
regularization of the brane gives a result that agrees with the ones
presented in the literature.

In order to directly compare with the results in the literature
we consider a pure AdS$_5$ background, a
boundary localized Higgs 
and a constant bulk
mass for the fermions $M=c/L_0$.
Following~\cite{Azatov:2009na} we define the Yukawa Lagrangian for a
boundary Higgs (for simplicity we focus on the down sector)
\begin{equation}
S_{\mathrm{Yuk}}=-\int d^4x dz \delta(z-L_1) a^3 
\Big[ 
\bar{Q}_L Y^{5D}_1 L_0 \phi D_R 
+\bar{Q}_R Y^{5D}_2 L_0 \phi D_L + \mathrm{h.c.}
\Big]+\ldots~,
\end{equation}
where we have already canonically normalized the Higgs boson 
$\phi$ (so that its vev is $v\sim L_1^{-1}$) and we have included a
term involving boundary values of fields with Dirichlet boundary
conditions. The latter terms do not involve the zero modes and
therefore do not affect the values of the effective
$z$-dependent Yukawa couplings appearing in the currents, 
that read in our case
\begin{eqnarray}
\lambda_{Dq}(z)&=&\delta(z-L_1) a_{M_D}^{-\frac{1}{2}}
(Y^{5D}_1)^\dagger L_1 a_{M_Q}^{\frac{1}{2}} f^Q_{q_L}, 
\label{lambdaDq}\\
\lambda_{Qd}(z)&=&\delta(z-L_1) a_{M_Q}^{\frac{1}{2}}
Y^{5D}_1 L_1 a_{M_D}^{-\frac{1}{2}} f^D_{d_R}. 
\end{eqnarray}
However, both terms contribute in principle to the $\Op_\varphi^{bc}$
term, which reads in our case
\begin{equation}
\Op_\varphi^{DQ}(z)=-\delta(z-L_1)L_1 \Big[
a_{M_D}^{-\frac{1}{2}}  Y^{5D\,\dagger}_1 a_{M_Q}^{\frac{1}{2}} \mathcal{P}_L
+a_{M_D}^{\frac{1}{2}}  Y^{5D\,\dagger}_2 a_{M_Q}^{-\frac{1}{2}}
\mathcal{P}_R
\Big]. \label{opvarphi}
\end{equation}
In fact, due to the chiral structure of the coefficient
$\gamma^{DQ;RL}_{q^i_L d^j_R}$ only the term proportional to $Y_2$
can give a non-vanishing contribution.
Now that we have the explicit expression for the relevant overlaps
with the Higgs, we only need the corresponding propagators to compute
the modified Yukawa couplings.
The $\beta$ coefficients are given by
\begin{eqnarray}
\beta^{D;R}_{q^i_Lq^j_L}&=&
-L_1^3
f^{Q\,\dagger}_{q_L}a_{M_Q}^{\frac{1}{2}} 
Y^{5D}_1 a_{M_D}^{-\frac{1}{2}}
\hat{P}^{(0)[--]}_{RR}(c_D,L_1,L_1)
a_{M_D}^{-\frac{1}{2}}
(Y^{5D}_1)^\dagger  a_{M_Q}^{\frac{1}{2}} f^Q_{q_L}, 
\\
\beta^{Q;L}_{d^i_R d^j_R}&=&
-L_1^3
f^{D\,\dagger}_{d_R}
a_{M_D}^{-\frac{1}{2}} 
Y^{5D\,\dagger}_1
a_{M_Q}^{\frac{1}{2}}
\hat{P}^{(0)[++]}_{LL}(c_Q,L_1,L_1)
a_{M_Q}^{\frac{1}{2}}
Y^{5D}_1 
a_{M_D}^{-\frac{1}{2}} 
f^D_{d_R},
\end{eqnarray}
where $a_M(L_1)=(L_0/L_1)^{2ML_0}$, $c_Q=M_QL_0$, $c_D=M_D L_0$  
and the
explicit expressions of the boundary propagators for the AdS case is
\begin{eqnarray}
&&\hat{P}_{LL}^{(0)[++]}(c,L_1,L_1)=
\hat{P}_{RR}^{(0)[--]}(-c,L_1,L_1)=
\frac{L_0^2}{(4c^2-4c-3)[1-(L_0/L_1)^{2c-1}]^2} \times
 \\
&&\qquad
\left[
(4c^2-4c-3) \left(\frac{L_0}{L_1}\right)^{-2}
-(2c-1)^2
+(3-2c) \left(\frac{L_0}{L_1}\right)^{-2c-1}
+(1+2c) \left(\frac{L_0}{L_1}\right)^{2c-3}
\right]. \nonumber 
\end{eqnarray}
See Eqs. (\ref{P:explicit:first}) and (\ref{P:explicit:rr}). The $\gamma$ coefficient on the other hand, involves propagators that
present discontinuities at the branes
\begin{equation}
\hat{P}^{(0)[++]}_{RL}(z,z^\prime) 
= \theta(z-z^\prime)
-\frac{1}{L_M}\int_{L_0}^z dz_1 a_M(z_1),
\end{equation}
and
\begin{equation}
\hat{P}^{(0)[--]}_{RL}(z,z^\prime) 
= -\theta(z^\prime-z)+\frac{1}{L_{-M}}
\int_{L_0}^{z^\prime} dz_1 a_{-M}(z_1).
\end{equation}
This discontinuities indicate that the coefficients $\gamma$ are
ambiguous in the thin brane limit and therefore the Dirichlet brane 
Yukawa term, proportional to $Y_2$, can potentially give a non-vanishing
contribution. In order to obtain a finite result we will replace the
brane terms in Eqs. (\ref{lambdaDq}-\ref{opvarphi}) 
with regulated delta functions
\begin{equation}
\delta(z-L_1)\to \delta_\epsilon(z-L_1)=\left\{
\begin{array}{l}
0,\quad z< L_1-\epsilon, \\
\frac{1}{\epsilon},\quad L_1-\epsilon\leq z \leq L_1.
\end{array}
\right.
\end{equation} 
Once the branes are regulated, we can perform all the relevant
integrals and then take the limit $\epsilon\to 0$. The (non-vanishing)
result we obtain is
\begin{eqnarray}
\gamma^{DQ;RL}_{q^i_Ld^j_R}&=&
\frac{L_1^3}{3} f^{Q\,\dagger}_{q_L} a^{\frac{1}{2}}_{M_Q} Y^{5D}_1
Y^{5D\,\dagger}_2
Y^{5D}_1 a_{M_D}^{-\frac{1}{2}} f^D_{d_R}. \label{gamma}
\end{eqnarray}
Using the fact that (see Appendix~\ref{trilinear:App})
\begin{equation}
m^{d(\mathrm{phys.})}_i \delta_{ij}- \frac{v}{\sqrt{2}} Y^d_{ij}
=\frac{v^3}{\sqrt{2}} \frac{(\alpha_{d\phi})_{ij}}{\Lambda^2},
\end{equation}
the result for $(\alpha_{d\phi})_{ij}$ deduced from
Eqs.(\ref{alphadphi},\ref{gamma})
exactly reproduces the results of Eq. (73) in
Ref.~\cite{Azatov:2009na}, taking into account the following dictionary between our notation and theirs:
\begin{eqnarray}
	v^{\mathrm{here}}=\sqrt{2}v^{[15]},\quad (a_{M_Q}^{\frac{1}{2}}f_{q_L}^Q)^{\mathrm{here}}=(\hat{F}_q)^{[15]},\quad (a_{M_D}^{-\frac{1}{2}}f_{d_R}^D)^{\mathrm{here}}=(\hat{F}_d)^{[15]}.
\end{eqnarray}
In Ref. \cite{Azatov:2009na} the
correct result was obtained, either by taking the
boundary limit of a bulk Higgs or by summing the effect of a number of
KK modes of the order of the inverse width of the regulated brane in
the case of a brane Higgs. In our case, this resummation is done
automatically by means of the 5D propagators which immediately shows
the ambiguity present in terms of Dirichlet Yukawa coupligns and the
need for a regularization of such couplings.

\section{Conclusions \label{conclusions}}

Models with warped extra dimensions represent calculable examples of
models in which EWSB proceeds through a strongly coupled conformal
sector. With the LHC currently probing the TeV scale, it is important
to have a handy way of computing the low energy effects of the largest
possible number of different models of strong EWSB. The
effective Lagrangian of models with warped extra dimensions arising
from the tree level integration of bulk gauge bosons has been know
for general background metrics and Higgs profiles for some time now. 
In this work we
have completed the most relevant part of the tree level (dimension 6)
effective Lagrangian by including the effect of an arbitrary number of
bulk fermions, again with general warp factor and Higgs profile. Our
general calculation involves the integration of the full 5D fields
(with the zero modes subtracted when present) by means of the 5D
propagators. This has the advantage of trading eigenvalue searches and
sums with integrals that can be easily done numerically. Also, it
makes apparent the subtleties related to brane localized terms involving
fields with Dirichlet boundary conditions as the automatic resummation
of all the modes gives a non-vanishing propagator near the brane for
certain components of fields with Dirichlet boundary conditions. 

Our results are completely general and can be applied to any model
with warped (or flat) extra dimensions, with or without an IR
brane~\cite{fermionsinsw}. 
This finally permits the complete calculation of the low energy
effects of a large variety of holographic models of strong EWSB,
including all the relevant electroweak and flavor effects. 

\section*{Note Added:} Upon completion of this work,
Ref.~\cite{Cabrer:2011qb} 
appeared in the arXives. In that reference, the contribution of bulk
fermions to the coupling of the down sector to the $Z$ boson is
computed using similar techniques to the ones presented here.

\acknowledgments{}

We would like to thank Manuel P\'erez-Victoria for useful
discussions. This work has been funded by MICINN (FPA2006-05294,
FPA2010-17915 and FPU programs) and Junta de Andaluc\'{\i}a (FQM 101,
FQM 03048 and FQM-6552).
\appendix

\section{Derivation of the fermionic Green's functions \label{green:App}}

Assuming no boundary terms, the quadratic operator in mixed
momentum/position space reads
\begin{equation}
\Op_K =
\Big[ \cancel{p} a_{M}^{-\gamma^5}  +  \gamma^5 \partial_z \Big].
\end{equation}
The different components of the fermionic Green's functions 
satisfy the following coupled equations
\begin{eqnarray}
a_M^{-1} \hat{P}^{RL} -L_1 \partial_z \hat{P}^{LL} &=& 0,
\qquad
a_M L_1 p^2 \hat{P}^{LL}+ \partial_z \hat{P}^{RL} = \delta(z-z^\prime),
\label{LL,RL}\\
a_M \hat{P}^{LR} +L_1 \partial_z \hat{P}^{RR} &=& 0,
\qquad 
a_M^{-1}L_1 p^2 \hat{P}^{RR}- \partial_z \hat{P}^{LR} = \delta(z-z^\prime),
\label{RR,LR}
\end{eqnarray}
and can be written in terms of the KK mode profiles as follows
\begin{eqnarray}
L_1 \hat{P}_p^{LL}(z,z^\prime)
&=&
\sum_n \frac{f_n^L(z) f_n^L(z^\prime)}{p^2-m_n^2},\label{fLfL}
\\
L_1 \hat{P}_p^{RR}(z,z^\prime)
&=&
\sum_n \frac{f_n^R(z) f_n^R(z^\prime)}{p^2-m_n^2},\label{fRfR}\\
\hat{P}_p^{RL}(z,z^\prime) &=& \hat{P}_p^{LR}(z^\prime,z) = 
\sum_n m_n   \frac{f_n^R(z) f_n^L(z^\prime)}{p^2-m_n^2}.\label{fLfR}
\end{eqnarray}
We can obtain equations for the $LL$ and $RR$ components by
iteration 
\begin{eqnarray}
L_1\Big[\partial_z a_M \partial_z+ a_M p^2\Big]
\hat{P}_p^{LL}(z,z^\prime)&=&\delta(z-z^\prime), \label{eq:PLL}
\\
L_1\Big[\partial_z a_{-M} \partial_z+ a_{-M} p^2\Big]
\hat{P}_p^{RR}(z,z^\prime)&=&\delta(z-z^\prime).\label{eq:PRR}
\end{eqnarray}
These equations are identical to the ones of a gauge boson with a
generalized metric $a_{\pm M}(z)$ for LH and RH components,
respectively. The zero momentum propagators for gauge bosons with an
arbitrary warp factor can be computed in a number of ways. The most
direct method is probably the one suggested recently
in~\cite{Cabrer:2010si} which amounts to solving directly the
equations that the Green's function \textit{at zero momentum} satisfy.
The resulting Green's functions at zero momentum, 
with zero modes subtracted and for an arbitrary warp factor $a$, read
(see~\cite{Cabrer:2010si})
\begin{eqnarray}
L_1 \hat{P}^{(0)[++]}
(a,z,z^\prime),&=&
-\frac{1}{L} \int_{L_0}^{z_<}dz_2 a^{-1}(z_2)\int_{L_0}^{z_2}dz_1
a(z_1)
-\frac{1}{L} \int_{z_>}^{L_1}dz_2 a^{-1}(z_2)\int_{z_2}^{L_1}dz_1
a(z_1)
\nonumber \\
&+&
\frac{1}{L^2} \int_{L_0}^{L_1}dz_1 a^{-1}(z_1) 
\int_{L_0}^{z_1} dz_2 a(z_2)
\int_{z_1}^{L_1} dz_3 a(z_3),
\\
L_1 \hat{P}^{(0)[--]}
(a,z,z^\prime)&=&
-\frac{
\int_{L_0}^{z_<} dz_1 a^{-1}(z_1)
~\int_{z_>}^{L_1} dz_2 a^{-1}(z_2)
}{
\int_{L_0}^{L_1} dz_3 a^{-1}(z_3)
},
\\
L_1 \hat{P}^{(0)[-+]}
(a,z,z^\prime)&=&
-\int_{L_0}^{z_<} dz_1 a^{-1}(z_1),
\\
L_1 \hat{P}^{(0)[+-]}
(a,z,z^\prime)&=&
-\int_{z_>}^{L_1} dz_1 a^{-1}(z_1),
\end{eqnarray}
where the superscript denotes the boundary conditions with a $+$
($-$) denoting Neumann (Dirichlet) boundary conditions and the first
(second) sign refering to the UV (IR) boundary and we have defined 
\begin{equation}
L\equiv \int_{L_0}^{L_1}dz\, a(z).
\end{equation}
The fermionic Green's functions $\hat{P}^{(0)}_{LL}$ and 
$\hat{P}^{(0)}_{RR}$ can be expressed in terms of these with the
replacement $a\to a_{\pm M}$ (and the understanding that the boundary
conditions correspond to the LH chirality of the 5D field). 
The $LR$ and $RL$ components can then be
trivially obtained from these by using the first equations in
(\ref{LL,RL}) or (\ref{RR,LR}). We can also use
$\hat{P}_{LR}(z,z^\prime)=\hat{P}_{RL}(z^\prime,z)$.

Let us now give the result for the different possibilities of boundary
conditions. 

\begin{description}
\item{- $[++]$ case} 

The LL and RR components are given by
\begin{eqnarray}
\hat{P}_{LL}^{(0)}(z,z^\prime)&=&
\hat{P}^{(0)[++]}(a_M,z,z^\prime), \label{P:explicit:first} \\
\hat{P}_{RR}^{(0)}(z,z^\prime)&=& \hat{P}^{(0)[--]}(a_{-M},z,z^\prime), 
\end{eqnarray}
whereas the mixed ones read
\begin{eqnarray}
\hat{P}_{LR}^{(0)}(z,z^\prime)&=&
\theta(z^\prime-z)-\frac{1}{L_M}\int_{L_0}^{z^\prime}dz_1 a_M(z_1),
\\
\hat{P}_{RL}^{(0)}(z,z^\prime)&=&
\theta(z-z^\prime)-\frac{1}{L_M}\int_{L_0}^{z}dz_1 a_M(z_1),
\end{eqnarray}

\item{- $[-+]$ case}

\begin{eqnarray}
\hat{P}_{LL}^{(0)}(z,z^\prime)&=&
\hat{P}^{(0)[-+]}(a_M,z,z^\prime), \\
\hat{P}_{RR}^{(0)}(z,z^\prime)&=&
\hat{P}^{(0)[+-]}(a_{-M},z,z^\prime), \\
\hat{P}_{LR}^{(0)}(z,z^\prime)&=&
-\theta(z-z^\prime), \\
\hat{P}_{RL}^{(0)}(z,z^\prime)&=&
-\theta(z^\prime-z), 
\end{eqnarray}

\item{- $[--]$ case}

\begin{eqnarray}
\hat{P}_{LL}^{(0)}(z,z^\prime)&=&
\hat{P}^{(0)[--]}(a_M,z,z^\prime), \\
\hat{P}_{RR}^{(0)}(z,z^\prime)&=&
\hat{P}^{(0)[++]}(a_{-M},z,z^\prime), \label{P:explicit:rr}\\
\hat{P}_{LR}^{(0)}(z,z^\prime)&=&
-\theta(z-z^\prime)+\frac{1}{L_{-M}}\int_{L_0}^z dz_1 a_M^{-1}(z_1), \\
\hat{P}_{RL}^{(0)}(z,z^\prime)&=&
-\theta(z^\prime-z)+\frac{1}{L_{-M}}\int_{L_0}^{z^\prime} dz_1 a_M^{-1}(z_1), 
\end{eqnarray}

\item{- $[+-]$ case}

\begin{eqnarray}
\hat{P}_{LL}^{(0)}(z,z^\prime)&=&
\hat{P}^{(0)[+-]}(a_M,z,z^\prime), \\
\hat{P}_{RR}^{(0)}(z,z^\prime)&=&
\hat{P}^{(0)[-+]}(a_{-M},z,z^\prime), \\
\hat{P}_{LR}^{(0)}(z,z^\prime)&=&
\theta(z^\prime-z), \\
\hat{P}_{RL}^{(0)}(z,z^\prime)&=&
\theta(z-z^\prime). \label{P:explicit:last}
\end{eqnarray}

\end{description}

\section{Trilinear couplings in the physical basis\label{trilinear:App}}

In this appendix we reproduce the couplings of the SM quarks to the SM
gauge bosons and Higgs with the Lagrangian in Eq. (\ref{Leff}) that
were computed in~\cite{delAguila:2000aa}. The masses of the SM quarks
in the physical basis read
\begin{eqnarray}
m^{u (\mathrm{phys.})}_i&=&
\frac{v}{\sqrt{2}}\left( \lambda^u_i 
- \frac{1}{4}[(V\alpha_{u\phi})_{ii} +(V\alpha_{u\phi})^\dagger_{ii}]
\frac{v^2}{\Lambda^2}\right), \\
m^{d (\mathrm{phys.})}_i&=&
\frac{v}{\sqrt{2}}\left( \lambda^d_i 
- \frac{1}{4}[(\alpha_{d\phi})_{ii} +(\alpha_{d\phi})^\dagger_{ii}]
\frac{v^2}{\Lambda^2}\right).  
\end{eqnarray}
In this basis, the SM quark couplings to the SM gauge and Higgs bosons
can be written as

\begin{eqnarray}
\lag^Z&=&-\frac{g}{2\cos \theta_W} \left( 
\bar{u}^i_L X^{uL}_{ij}
\gamma^\mu u^j_L+
\bar{u}^i_R X^{uR}_{ij}
\gamma^\mu u^j_R \right.\nonumber \\
&&\left.-\bar{d}^i_L X^{dL}_{ij}
\gamma^\mu d^j_L-
\bar{d}^i_R X^{dR}_{ij}
\gamma^\mu d^j_R - 2\sin^2\theta_W J^\mu_{\mathrm{EM}}\right) Z_\mu,
\nonumber \\ 
\lag^W&=&-\frac{g}{\sqrt{2}}(
\bar{u}^i_L W^{L}_{ij}
\gamma^\mu d^j_L+
\bar{u}^i_R W^{R}_{ij}
\gamma^\mu d^j_R)W^+_\mu +\textit{h.c.}
, \label{lag:ZWH}\\
\lag^H&=&-\frac{1}{\sqrt{2}}(
\bar{u}^i_L Y^{u}_{ij}
 u^j_R+
\bar{d}^i_L Y^{d}_{ij} d^j_R + \mathit{h.c.} ) H
\nonumber \\ 
&&+ \left( 
\bar{u}^i_L Z^{uL}_{ij}
\gamma^\mu u^j_L+
\bar{u}^i_R Z^{uR}_{ij}
\gamma^\mu u^j_R-\bar{d}^i_L Z^{dL}_{ij}
\gamma^\mu d^j_L-
\bar{d}^i_R Z^{dR}_{ij}
\gamma^\mu d^j_R \right) i\partial_\mu H. \nonumber 
\end{eqnarray}
The unbroken $U(1)_Q$ protects the terms proportional to
$J^\mu_{\mathrm{EM}}$. 
The expressions to order $1/\Lambda^2$ of the coupling matrices 
$X$, $W$, $Y$ and $Z$ in terms of the coefficients $\alpha_x$ are:
\begin{eqnarray}
X^{uL}_{ij}&=& \delta_{ij}-\frac{1}{2}\frac{v^2}{\Lambda^2}
V_{ik}(\alpha^{(1)}_{\phi q}+\alpha^{(1)\dagger}_{\phi q}
-\alpha^{(3)}_{\phi q}-\alpha^{(3)\dagger}_{\phi q}
)_{kl}V^\dagger_{lj}, \nonumber \\
X^{uR}_{ij}&=& -\frac{1}{2}\frac{v^2}{\Lambda^2}
(\alpha_{\phi u}+\alpha_{\phi u}^\dagger
)_{ij}, \nonumber \\
X^{dL}_{ij}&=& \delta_{ij}+\frac{1}{2}\frac{v^2}{\Lambda^2}
(\alpha^{(1)}_{\phi q}+\alpha^{(1)\dagger}_{\phi q}+
\alpha^{(3)}_{\phi q}+  
\alpha^{(3)\dagger}_{\phi
q})_{ij}, \nonumber \\
X^{dR}_{ij}&=& \frac{1}{2}\frac{v^2}{\Lambda^2}
(\alpha_{\phi d}+\alpha_{\phi d}^\dagger
)_{ij},  \label{couplings}\\
W^{L}_{ij}&=& \tilde{V}_{ik}\left(\delta_{kj}+\frac{v^2}{\Lambda^2}
(\alpha^{(3)}_{\phi q})_{kj}\right), \nonumber \\
W^{R}_{ij}&=& -\frac{1}{2}\frac{v^2}{\Lambda^2}
(\alpha_{\phi \phi})_{ij}, \nonumber \\
Y^{u}_{ij}&=& \delta_{ij}\lambda^u_j - \frac{v^2}{\Lambda^2} 
\left(V_{ik}(\alpha_{u \phi})_{kj} +\frac{1}{4} \delta_{ij}
[V_{ik} (\alpha_{u \phi})_{kj} + (\alpha_{u \phi}^\dagger)_{ik}
V^\dagger_{kj}]\right), 
\nonumber \\ 
Y^{d}_{ij}&=& \delta_{ij}\lambda^d_j - \frac{v^2}{\Lambda^2} 
\left((\alpha_{d \phi})_{ij}  + \frac{1}{4} \delta_{ij} (\alpha_{d
\phi}+\alpha_{d\phi}^\dagger)_{ij} \right) , \nonumber \\ 
Z^{uL}_{ij}&=& -\frac{1}{2}\frac{v}{\Lambda^2}
V_{ik}(\alpha^{(1)}_{\phi q}-\alpha^{(1)\dagger}_{\phi q}
-\alpha^{(3)}_{\phi q}+\alpha^{(3)\dagger}_{\phi q}
)_{kl}V^\dagger_{lj}, \nonumber \\
Z^{uR}_{ij}&=& -\frac{1}{2}\frac{v}{\Lambda^2}
(\alpha_{\phi u}-\alpha_{\phi u}^\dagger
)_{ij}, \nonumber \\
Z^{dL}_{ij}&=& \frac{1}{2}\frac{v}{\Lambda^2}
(\alpha^{(1)}_{\phi q}-\alpha^{(1)\dagger}_{\phi q}+
\alpha^{(3)}_{\phi q}-  
\alpha^{(3)\dagger}_{\phi
q})_{ij}, \nonumber \\
Z^{dR}_{ij}&=& \frac{1}{2}\frac{v}{\Lambda^2}
(\alpha_{\phi d}-\alpha_{\phi d}^\dagger
)_{ij}.\nonumber
\end{eqnarray}
We have introduced the unitary matrix
\begin{equation}
\tilde{V} = V+ \frac{v^2}{\Lambda^2} (V A_L^d - A_L^u V) \,,
\end{equation}
with
\begin{eqnarray}
(A^u_L)_{ij} & = & \frac{1}{2}(1-\frac{1}{2} \delta_{ij})
\frac{\lambda^u_i (V 
\alpha_{u\phi})^\dagger_{ij} + (-1)^{\delta_{ij}}
(V\alpha_{u\phi})_{ij} \lambda^u_j} 
{(\lambda^u_i)^2-(-1)^{\delta_{ij}}(\lambda^u_j)^2} \, ,\nonumber \\
(A^d_L)_{ij} & = & \frac{1}{2}(1- \frac{1}{2} \delta_{ij})
\frac{\lambda^d_i 
(\alpha_{d\phi})^\dagger_{ij} + (-1)^{\delta_{ij}}
(\alpha_{d\phi})_{ij} \lambda^d_j} 
{(\lambda^d_i)^2-(-1)^{\delta_{ij}}(\lambda^d_j)^2} \, .\nonumber \\
\end{eqnarray}
Note that, to order $1/\Lambda^2$, we can substitute $V$ by
$\tilde{V}$ everywhere in Eq.~(\ref{couplings}), so that the different
couplings 
depend on only one unitary matrix.

\end{document}